\documentclass[
 reprint,
 amsmath,amssymb,
 aps,
]{revtex4-2}

\usepackage{graphicx}
\usepackage{float}
\usepackage{dcolumn}
\usepackage{bm}
\usepackage{mathtools}
\usepackage{physics}
\usepackage{hyperref}
\newcommand{\argmin}{\mathop{\mathrm{argmin}}\limits}
\newcommand{\act}{{\textrm{act}}}
\newcommand{\SE}{{\textrm{SE}}}
\newcommand{\DE}{{\textrm{DE}}}
\newcommand{\beginsupplement}{
    \setcounter{figure}{0}
    \setcounter{section}{0}
    \setcounter{equation}{0}
    \renewcommand{\thefigure}{S\arabic{figure}}
    \renewcommand{\theequation}{S\arabic{equation}}
    \renewcommand{\thefigure}{S\arabic{figure}}
    \renewcommand{\thetable}{S\arabic{table}}
    \renewcommand\thesection{\Alph{section}}
}

\begin{document}

\title{Improve it or lose it:\\evolvability costs of competition for expression}

\author{Jacob Moran}
\affiliation{
Department of Physics, Washington University in St. Louis, St. Louis, MO 63130
}
\author{Devon Finlay}
\affiliation{
Department of Physics, Washington University in St. Louis, St. Louis, MO 63130
}
\author{Mikhail Tikhonov}
\affiliation{
Department of Physics, Washington University in St. Louis, St. Louis, MO 63130
}
\date{December 16, 2020}
\begin{abstract}
    Expression level is known to be a strong determinant of a protein's rate of evolution. But the converse can also be true: evolutionary dynamics can affect expression levels of proteins. Having implications in both directions fosters the possibility of a feedback loop, where higher expressed systems are more likely to improve and be expressed even higher, while those that are expressed less are eventually lost to drift. Using a minimal model to study this in the context of a changing environment, we demonstrate that one unexpected consequence of such a feedback loop is that a slow switch to a new environment can allow genotypes to reach higher fitness sooner than a direct exposure to it.
\end{abstract}
\maketitle

Expression level of proteins can influence their evolution. For instance, substantial evidence suggests that lower-expressed proteins are less protected from drift, whereas highly expressed proteins are under stronger purifying selection \cite{drummond2005,savageau1977,hwa2009,zhang2015}. 

Conversely, evolution can also affect expression. In many cases, expression levels are determined by functional requirements; e.g.\ membrane synthesis must match the cells' growth rate. But for proteins or pathways that are dispensable or partially redundant, evolution can affect expression significantly. The most intuitive mechanism is through genetic drift: a protein disabled by a deleterious mutation becomes a metabolic burden (or may be directly toxic), favoring loss of expression.
	
Since partial redundancy is believed to be widespread \cite{pilpel2009}, this creates a theoretical possibility of a feedback loop. Consider an organism with several partially substitutable systems or pathways fulfilling a similar function; for example, several metabolic pathways to satisfy its requirement for carbon, or several sensing modalities to respond to environmental cues. It is plausible that the systems used more, being under a stronger selection pressure, would be more likely to improve and be used even more. In contrast, the lesser expressed systems could be more likely to deteriorate and be used even less. 

\begin{figure}[b]
    \includegraphics{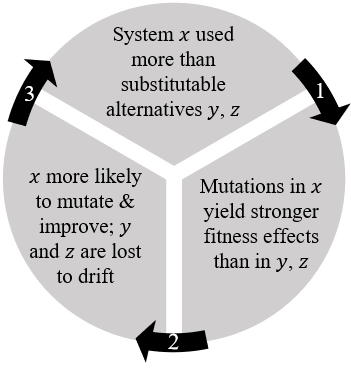}
    \caption{\textbf{The ``improve it or lose it'' feedback loop.} In this schematic, $x$, $y$, and $z$ are partially substitutable systems fulfilling a similar function (e.g., metabolic pathways for alternative sources of carbon). Adaptive mutations in the highest-used system $x$ have stronger fitness effects than $y$, $z$ (arrow 1). The stronger selection pressure makes system $x$ more likely to  mutate and improve (arrow 2). This improvement in $x$ allows the organism to rely on it even more (arrow 3), completing the loop.}
    \label{fig:1}
\end{figure}

Here, we use a minimal theoretical model \cite{tikhonov2020} to exhibit this feedback and examine its consequences. This process -- effectively a ``competition for expression'' -- could be viewed as an extension of Savageau's ``use it or lose it'' principle (Fig.~\ref{fig:1}), and is conceptually similar to the generalist-to-specialist transition of ecological specialization \cite{lenski2000,kawecki1997}, where a population increasingly improves its use of one environmental resource at the expense of others.  At the same time, this expectation would seem to be in conflict with the findings of studies such as Ref.~\cite{drummond2005}, which found that the highly expressed proteins evolve slower rather than faster. We reconcile this apparent conflict by showing that, at least in our model, the correlation between expression level and evolutionary rate changes sign as adaptation proceeds: the ``improve it or lose it'' feedback pertains to early stages dominated by adaptive mutations. Our argument suggests that following a drastic environmental change, the correlation between expression and evolutionary rate may transiently invert. Finally, we demonstrate a curious consequence of this feedback loop for adaptation to novel environments: namely, that a gradual change to a new environment can lead to higher fitness faster than direct exposure.

\section{\label{sec:level1}Model and Context}

\begin{figure}[b]
    \includegraphics{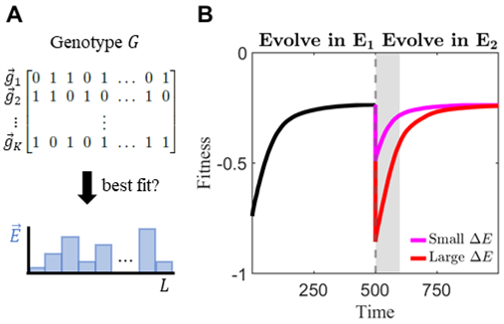}
    \caption{\textbf{A context to study the ``improve it or lose it'' feedback loop}. (A) In the toolbox model, a genotype is a matrix representing the available ``systems'' an organism can (linearly) combine to approximate the optimal phenotype required by the environment, $\vec{E}$. The coefficients of the best approximation are interpreted as usage levels $a_\mu$, serving as a proxy for expression. Matrix elements are chosen to be binary ($0$ or $1$) so that mutations in the evolutionary process can be implemented as bit flips. (B) Fitness trajectories of initially random genotypes evolving under $\vec{E}_1$ before switching to $\vec{E}_2$ a distance $\Delta E$ away. We choose to study the feedback loop and its consequences during the early-time dynamics after switching (gray region). } 
    \label{fig:2}
\end{figure}

To study the ``improve it or lose it'' feedback loop, we need an evolutionary model that explicitly includes a notion of usage/expression. For this reason, we adopt the toolbox model from Ref.~\cite{tikhonov2020}, summarized in Fig.~\ref{fig:2}A.

Briefly, we think of a genotype as encoding a set of $K$ systems that can be used at different levels to optimize the fitness of the organism in a given environment. Mathematically, we represent the $K$ systems as basis vectors $\{\vec g_\mu\}$ ($\mu=1\dots K$) and the environment as a target vector $\vec{E}$ in an abstract $L$-dimensional space (which can be interpreted as the phenotype space~\cite{tikhonov2020}). The fitting problem can be written as,
\begin{equation}\label{eq:Usage}
    \{a_\mu\} = \argmin_{a_\mu\geq 0} \norm{\vec{E} - \sum_\mu a_\mu \vec{g}_\mu},
\end{equation}
where the environment-dependent coefficients $\{a_\mu\}$ can be interpreted as the extent to which the organism relies on a given system $\vec g_\mu$ in $\vec E$. The quality of fit, which these $\{a_\mu\}$ optimize, can then be interpreted as the fitness of the genotype $G=\{\vec g_\mu\}$ in environment $\vec E$:
\begin{equation}\label{eq:Fitness}
F(\mathrm{G},\vec{E}) = - \min_{a_\mu\geq 0} \norm{\vec{E} - \sum_\mu a_\mu \vec{g}_\mu}.
\end{equation}
In Ref.~\cite{tikhonov2020}, the coefficients $\{a_\mu\}$ are called ``expression levels''; however, conceptually, they correspond more closely to the intuitive notion of ``usage''. Indeed, a larger $a_\mu$ in this model corresponds to a system whose deletion would have a stronger fitness effect, rather than one present in a larger copy number (although in practice, the two properties are, of course, correlated~\cite{he2005}). Throughout this work, we refer to $\{a_\mu\}$ as usage coefficients.

To simulate the evolutionary process, we assume the regime of strong selection and rare mutations so that we need only track the evolutionary trajectory of a single genotype \cite{gillespie1982} (see \textit{SI}). Fig. \ref{fig:2}B shows an example of fitness dynamics of random initial genotypes first exposed to a random environment $\vec{E}_1$ and then to a different random environment $\vec{E}_2$. The feedback loop we will describe is already present during the early-time dynamics of evolution in $\vec{E}_1$; however, we choose to focus on the time period that follows the environment switch (shaded gray region). This will allow us to use the difference between the two environments, $\Delta E = \norm{\vec{E}_2-\vec{E_1}}$ as a natural control parameter (see \textit{SI} for parameterization of environment pairs $(\vec{E}_1,\vec{E}_2)$). 

In what follows, we use $\vec{E}$ vectors of unit length so that fitness is constrained to $-1 \leq F \leq 0$. We fix $L=40$ and vary $K$, and consider genotype matrices with binary values, 0 or 1, initialized randomly with probability $p=0.5$ of being 1. Since environments are represented by unit vectors with positive components, $\Delta E$ is confined to the range $\Delta E \in \left[0,\sqrt{2}\right]$. We will show that $\Delta E$ controls the strength of the feedback loop, with stronger  changes in environment (large $\Delta E$) inducing stronger feedback.

\section{The toolbox model exhibits the ``Improve it or lose it'' feedback}
\begin{figure*}
    \centering
    \includegraphics[width=1\textwidth]{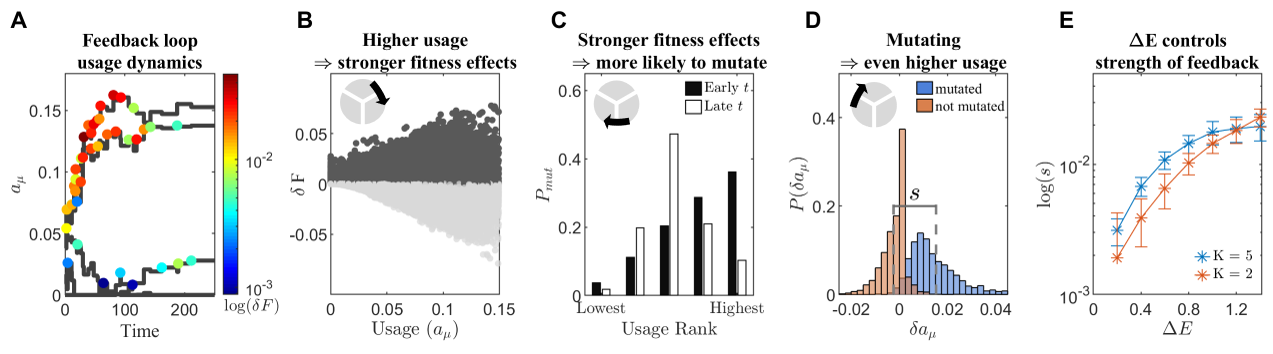}
    \caption{\textbf{The toolbox model exhibits the feedback loop.}  (A) An example of evolutionary dynamics of usage coefficients after a genotype adapted to a random environment $E_1$ is switched to another random environment $E_2$ with $\Delta E \equiv |\vec E_1-\vec E_2| = 1$. Despite similar usage initially, by $t=100$ only two of $K=5$ systems remain in use. Dots mark the systems in which a beneficial mutation arose, color indicates fitness effect (red is strongest). In panels (B-D), we examine the statistics of usage dynamics and mutation effects within the first 3 time steps of 20 trajectories in 15 random environment pairs with the same $\Delta E=1$. Inset pictograms refer to feedback steps as shown in Fig.~\ref{fig:1}. (B) Fitness effects of all available mutations in each system versus system usage. Dark and light gray points are beneficial and neutral/deleterious mutations, respectively. Higher-used systems possess stronger fitness effects. (C) Probability of a system to mutate, plotted against its usage rank (ascending order). At early times (black bars), higher used systems are more likely to mutate. As the strong beneficial mutations in highest-used systems are depleted, the probability of mutating shifts towards lower used systems (white bars). (D) Distribution of change in usage of a system that just mutated (blue), or a system that failed to mutate (red). The difference in means of these conditional probability distributions, $s$, quantifies the strength of the feedback loop. (E) The strength of the feedback loop $s$ is controlled by the magnitude of environmental change $\Delta E$. Error bars represent 1 standard deviation (SD) over 300 replicates.}
    \label{fig:3}
\end{figure*}

Fig.~\ref{fig:3}A depicts a representative trajectory of the ``improve it or lose it'' feedback realized in the toolbox model. The panel shows the dynamics of usage coefficients after a genotype with $K = 5$ systems, pre-adapted to some environment $\vec E_1$, was switched to a different environment $\vec E_2$, with $\Delta E=1$ (random environment pairs with a given $\Delta E$ were generated as described in the \textit{SI}). Note that, after each mutation, the usage coefficients are re-optimized according to eq.~\ref{eq:Usage} and thus change discontinuously (see \textit{SI} and Ref.~\cite{tikhonov2020}); however, these steps are typically small, creating an illusion of smooth dynamics. 
We see that strong adaptive mutations initially concentrate in the two systems with highest usage (frequent redder dots). As they mutate, they also rise in usage, $a_\mu$. In contrast, the lower-used systems decrease in usage, and mutate only rarely, with relatively weak fitness effects (bluer dots). 

Although the details of these dynamics are shaped by Eq.~\ref{eq:Fitness} and are of course model-dependent, on a qualitative level the instability driving a subset of usage coefficients up at the expense of others can be directly traced to the feedback loop summarized in Fig. 1, as we will now show.

First, agreeing with the intuitive notion of $\{a_\mu\}$ as ``usage'', systems with higher $a_\mu$ tend to harbor stronger fitness effects. To see this in our model, we plot the fitness effects of all available mutations against the usage coefficient of the system where they occur (Fig.~\ref{fig:3}B). As expected, both beneficial (dark gray) and deleterious (light gray) mutations are stronger in systems that have a higher usage coefficient $a_\mu$.

As a result, higher-used systems are more likely to mutate, because mutations with a larger fitness effect are more likely to escape drift and fix in the population \cite{kimura1962}. The black bars in Fig.~\ref{fig:3}C show the early-time probability of each system to mutate, plotted against its usage rank.

Finally, after a system mutates, its usage typically increases (Fig.~\ref{fig:3}D, blue). In contrast, systems that do not mutate typically drop in usage (Fig.~\ref{fig:3}D, red). In our model, this also is ultimately a consequence of Eq.~(2), but it is not the model that justifies this behavior. Rather, it is this behavior that justifies using the model, making it appropriate for studying the feedback loop this biologically plausible feature induces. In summary, Fig.~\ref{fig:3}B-D demonstrates all three arrows from Fig.~\ref{fig:1} at play in our model.

Since a greater separation between the distributions of Fig.~\ref{fig:3}D would entail stronger feedback, we can use the difference in the mean of these conditional distributions, denoted as~$s$, as a measure of the feedback strength. Fig.~\ref{fig:3}E demonstrates that, as expected, the feedback becomes stronger (increasing $\log s$) as the change in environment becomes more severe.

The rapid evolution of highly used systems (Fig.~\ref{fig:3}C) may seem to be at odds with experimental work showing that highly expressed proteins evolve slowest~\cite{drummond2005}. However, the mechanism described here is fully compatible with the explanations previously proposed for this experimental result. The effect shown in Fig.~\ref{fig:3}B (higher used systems have stronger fitness effects) applies to both beneficial and deleterious mutations. For early stages of adaptation driven by beneficial mutations (as considered here), this means the most-used systems will evolve first. However, at later stages, as beneficial mutations are depleted, the same argument dictates that the most-used systems become the most protected, and evolve slowest. We illustrate this effect by replotting the per-system mutation probabilities at a later time (Fig.~\ref{fig:3}C, white bars); the probability of mutating begins to shift from higher used to lesser used systems. Indeed, this transient flip in correlation between expression and evolutionary rate is consistent with recent analysis of evolutionary rates in yeast \cite{ascencio2017}.

\section{The cost to evolvability}
Intuitively, one might expect that the competition for usage mediated by the ``improve it or lose it'' feedback loop may be detrimental for the organism, since it effectively reduces the number of systems it has available. Implementing this effect in a simple model allows us to make this intuition precise. We will see that, at least in our model, the feedback loop exhibited above reduces the adaptive potential of the genotype, and mitigating its effects can allow for faster adaptation.

\begin{figure*}
    \centering
    \includegraphics[width=1\textwidth]{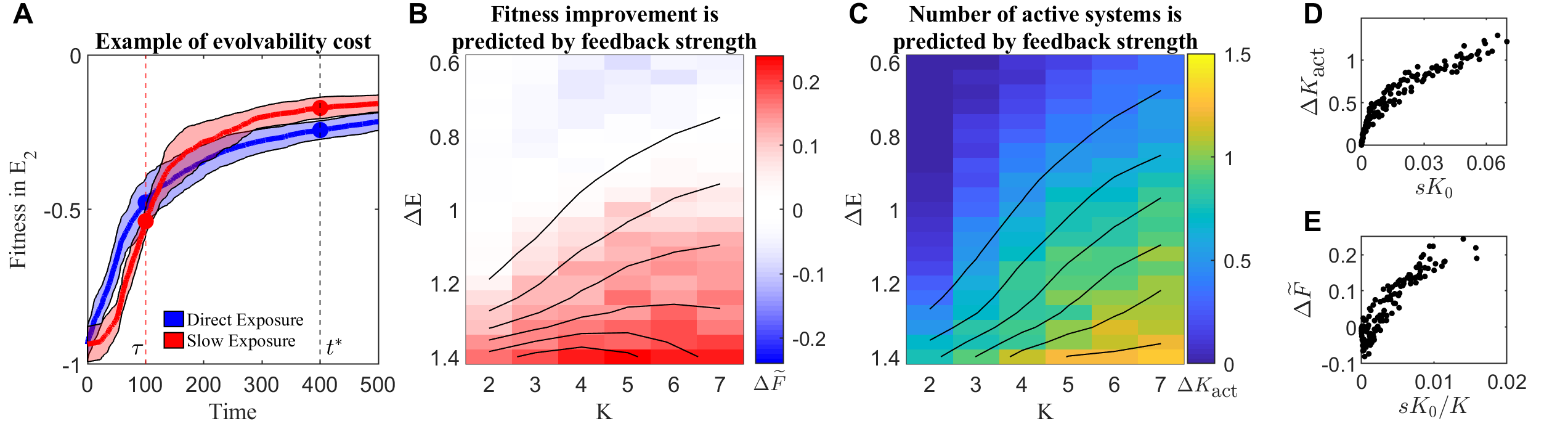}
    \caption{\textbf{Higher evolvability from slow exposure than direct exposure.} (A) Fitness, evaluated in environment $\vec{E}_2$, of genotypes that are either directly exposed (DE) to $\vec{E}_2$ at $t = 0$ (blue trace) or slowly exposed (SE) to $\vec{E}_2$ over a time $\tau = 100$ according to the protocol defined in Eq.~\eqref{eq:LinearRelaxProt} (red trace). Each trace shows mean $\pm 1$ SD (shading) of 20 replicate trajectories of genotypes with $K=4$ systems in a random environment pair with $\Delta E =1.4$. Colored dots highlight that slow exposure leads to higher long-term fitness, despite slower fitness gain initially. The relative improvement in fitness, $\Delta\widetilde{F}$, is measured at an arbitrarily late time point $t^* = 400$ (see \textit{SI} Fig.~S3C for later $t^*$). (B) Heatmap of the long-term relative fitness improvement, $\Delta\widetilde{F}$. Contour lines show $\Delta\widetilde{F}$ can be predicted by the feedback loop strength $s$ and the number of initially inactive systems $K_0$ (see panel E). Here and in the remaining panels, results are averages over 20 trajectories in 15 random environment pairs with varying $\Delta E$. (C) Heatmap of $\Delta K_{\act}$, the average difference in number of active systems ($a_\mu > 10^{-3}$) between the SE and DE protocols at $t=\tau$. Contour lines show it is predicted by the product $s K_0$; see panel D. (D) $\Delta K_{\act}$, the increased number of active systems at $t=\tau$, is predicted by $s K_0$, measured at trajectory start. (E) The long-term fitness improvement $\Delta\widetilde{F}$ at $t=t^*$ is predicted by $s K_0/K$, measured at trajectory start.}
    \label{fig:4}
\end{figure*}

For this, we compare the fitness trajectories of genotypes evolving in conditions that  exacerbate the feedback and those that weaken it. Specifically, starting from a genotype pre-adapted to $\vec E_1$, we compare two ways of adapting it to a new, strongly different environment $\vec E_2$: either by exposing it to $\vec E_2$ directly (as discussed above), or by changing the environment from $\vec E_1$ to $\vec E_2$ slowly (on a timescale that is slow compared to mutation fixation). By avoiding large environment jumps, we expect the gradual switch to weaken the feedback loop. The question we ask is which exposure protocol will ultimately lead to higher fitness in the environment of interest, $\vec E_2$.

An example of this comparison is shown in Fig.~\ref{fig:4}A. The red curve shows fitness (in the environment of interest $\vec E_2$) for genotypes evolving under the slow-exposure  protocol, implemented by linearly relaxing the environment vector from $\vec{E}_1$ to $\vec{E}_2$ over a time $\tau$:
\begin{equation}\label{eq:LinearRelaxProt}
\vec E(t) = \left\{
\begin{aligned}
&\mathrm{normalize}\left[\vec E_2+\frac{\tau-t}{\tau}(\vec E_1-\vec E_2)\right]&&\text{if }t<\tau\\
&\vec E_2 && \text{if }t\ge\tau
\end{aligned}\right.
\end{equation}
(the environment vector in our model is always normalized to unit length). The $\tau$ we use is large relative to the typical time between mutations ($\tau=100$; compare to  Fig.~\ref{fig:3}A). The red curve $F_{\SE}(t)$ (slow exposure) is to be compared to the blue curve $F_{\DE}(t)$ (direct exposure), showing fitness of the same initial genotypes evolving directly in $\vec E_2$.

The vertical dashed line at $t=\tau$ marks the timepoint where the ``red genotypes'' evolving under the slow-switching protocol are finally exposed to $\vec E_2$ for the first time. It is therefore not surprising that they are less fit than the ``blue genotypes'', who have been evolving in $\vec E_2$ from the start ($F_{\SE}(\tau)<F_{\DE}(\tau)$; red curve below the blue). However, while more fit, the blue genotypes are manifestly less evolvable: From $t=\tau$ onwards, both red and blue curves document evolution in the same environment $\vec E_2$, but the red curve gains fitness much faster, and overtakes the blue.

To quantify the strength of this effect, we consider the relative improvement of fitness provided by the smooth protocol, compared to direct exposure:
\begin{equation}\label{eq:EvolvabilityMetric}
    \Delta \widetilde{F}(t^*) \equiv \frac{F_{\SE}(t^*) - F_{\DE}(t^*)}{\abs{F_{\DE}(t^*)}}.
\end{equation}
While initially negative, in the example of Fig.~\ref{fig:4}A this quantity becomes positive at a later time. To demonstrate the robustness of this observation, Fig.~\ref{fig:4}B shows $\Delta \widetilde{F}(t^*)$ for a range of $K$ and $\Delta E$, computed at an arbitrary late timepoint $t^*=400$ (see \textit{SI} Fig.~S3C for $\Delta\widetilde{F}$ at a later value of $t^*$). We see that, at large $\Delta E$, the slow-switching protocol consistently outperforms direct exposure. While the scenario of an organism possessing $K=7$ competing systems fulfilling a similar function is arguably unrealistic, we note that the effect is already present at $K=2$. (For the purposes of illustration, the example in panel A used $K=4$ and $\Delta E=1.4$, when the effect is strongest.) Note that, for simplicity, in Fig.~\ref{fig:4}B our slow exposure protocol~\eqref{eq:LinearRelaxProt} used the same value of the relaxation time $\tau=100$ for all $K$ and $\Delta E$; optimizing over this parameter could of course render the effect stronger.

The origin of this effect is the ``improve it or lose it'' instability affecting the genotypes undergoing an abrupt environment switch, effectively leaving them with fewer systems. To confirm this, we record the average number $K_\act^{\DE}$, $K_\act^{\SE}$ of ``active'' systems (usage $a_\mu > 0.001$) observed at time $t = \tau$ under both protocols. As expected, a slow environment change leaves more systems active; the difference $\Delta K_\act\equiv K_\act^{\SE}-K_\act^{\DE}$ is shown in Fig.~\ref{fig:4}C and exhibits a trend similar to Fig.~\ref{fig:4}B. Since unused systems harbor weak mutations only (cf. Fig.~\ref{fig:3}B), a genotype with few active systems finds itself on a fitness plateau, and its rate of fitness gain is reduced.

Finally, we can quantitatively relate both effects to the strength of the feedback loop as defined above. To start, we focus on the increase in the number of active systems $\Delta K_\act$ in Fig.~\ref{fig:4}C. Denote $K_0$ the number of inactive systems at time $t=0$ (immediately after the environment switch; usage $a_\mu < 0.001$). This is the number of systems that the slow-exposure protocol could conceivably ``rescue''. One expects $\Delta K_\act$ to scale with $K_0$, and if our argument is correct, it should also scale with the strength of the feedback loop $s$. Indeed, we find $\Delta K_\act$ to be predicted by the product $s K_0$ (Fig.~\ref{fig:4}D). 
The availability of these additional systems translates into additional adaptive opportunities, and ultimately a higher fitness. In a strongly epistatic model like ours, the exact relationship to the long-term fitness is hard to predict. Nevertheless, it is reasonable to expect the fractional effect on fitness $\Delta \widetilde F$ to at least correlate with the fractional effect on the number of active systems $\Delta K_\act/K$. If so, then $\Delta \widetilde F$ should correlate with $s K_0/K$, an expectation confirmed in panel Fig.~\ref{fig:4}E. Given the approximate nature of this argument, the correlation observed in Fig.~\ref{fig:4}E is in fact surprisingly good. For convenience, the same $s K_0$ and $s K_0 / K$ data, Gaussian-smoothed for visualization purposes, are shown as contour lines superimposed on the heatmaps of Fig.~\ref{fig:4}B,~C. It is worth emphasizing that our definition of the feedback strength $s$ is computed from the statistics of the first 3 mutations, which only take $t \sim 7 \pm 5$ to occur; and $K_0$ is similarly measured at the very start of the trajectory. Nevertheless, at least in our model, these early-time properties are predictive of the long-term evolutionary outcome at $t^*=400$.
\section{Discussion}
In this work, we used a minimal model to
explore a possible feedback loop between the usage of a system and its rate of evolution. Within this model, we demonstrated that this feedback loop is particularly pronounced after strong shifts in the selecting environment and can negatively impact evolvability (future fitness gain). In particular, we described a mechanism by which a slow switch to a new environment can allow the genotypes to reach higher fitness sooner than a direct exposure to it.

A situation where exposure to a different environment $E'$ can help evolve better fitness in $E$ than a direct exposure to $E$ itself is not, in itself, novel. One well-established scenario for this to occur is the crossing of fitness valleys (or plateaus): much like an enzyme that catalyzes a reaction by stabilizing the reaction transition state, a transient exposure to $E'$ can facilitate reaching a higher fitness peak by enabling prerequisite mutations that would otherwise be unfavorable (or neutral) \cite{ostermeier2016,schwab2014}. However, the scenario described here is particularly interesting because the fitness plateau is not an idiosyncratic property of a particular landscape, but emerges through evolution itself. Fitness landscapes of evolved systems are themselves shaped by evolution \cite{desai2015,orr2003}, and at least in our model, the feedback mechanism we described generically induces a fitness plateau following an abrupt environmental change.

It is worth stressing that we considered beneficial mutations only. Clearly, if deleterious mutations were included, our feedback loop would become even stronger: in addition to the effect described, the lesser-used systems would also be less protected from drift \cite{allendorf1978,king1979,ohta1989}. This observation could be seen as the traditional manifestation of the ``use it or lose it'' principle; in particular, the problem of maintaining redundancy in the face of drift has been extensively discussed \cite{smith1997}. Focusing on beneficial mutations only, and thus explicitly excluding any drift-dependent effects, allows us to highlight a novel aspect. Unlike the discussion of Ref.~\cite{smith1997}, here, no system is ever fully redundant, and all remain under selection. Nevertheless, some are progressively lost even in the absence of deleterious mutations -- simply because the beneficial mutations preferentially target the systems used more, and those that fail to improve become obsolete. This mechanism is clearly analogous to the Red Queen effect \cite{valen1973}, except here it applies to an effective competition for expression. In this way, the loss of evolvability described in Fig.~\ref{fig:4} can be seen as a form of a conflict of levels of selection \cite{okasha2006}.

\textbf{Data availability:} All simulations were performed in MATLAB (Mathworks, Inc.). The associated code, data and scripts to reproduce all figures in this work are available at Mendeley Data, \href{http://dx.doi.org/10.17632/zdsnttv2dt.1}{http://dx.doi.org/10.17632/zdsnttv2dt.1} \cite{moran2020}.

\textbf{Acknowledgements:} We thank C.~Holmes, S.~Kaplan, S.~Kuehn and S.~Maslov for helpful discussions. 

\bibliography{evoElsewhereBibV1}




\clearpage
\onecolumngrid
\beginsupplement

\section*{Supplementary Information}
\section{\label{sec:level1}Modeling the Evolutionary Process}
\begin{figure}[H]
    \centering
    \includegraphics[width=8cm]{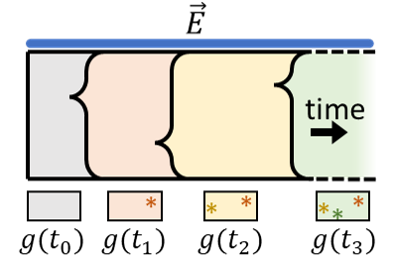}
    \caption{Evolution in strong selection, rare mutation regime.}
    \label{fig:S1}
\end{figure}

For simplicity in simulating the process of evolution, we work in the regime where mutations are rare and selection is strong. In doing so, we only need to track the evolutionary trajectory of a single genotype matrix, representing a clonal population that evolves by sequential mutations that sweep through the entire population as schematized in Fig.~\ref{fig:S1}. Computationally, we implement a Gillespie-style algorithm, where each loop iteration updates the genotype matrix by mutation and selection. Having chosen the genotype matrix to be binary, we can implement point mutations as simple bit-flips of one of the matrix elements. Within each step of the algorithm, all point mutations of the current genotype are enumerated. The fitness effect of each mutation relative to the current genotype fitness is computed using eq.~(2) of the main text (reoptimizing the usage coefficients for each mutant). From only the beneficial mutations, one is randomly selected to fix for the next iteration, with the probability of being selected weighted by fitness effect (strong selection, rare mutation regime; \cite{gillespie1982,kimura1962}). Finally, we also update the state of the environment within the Gillespie loop if the environment target vector is dynamic (e.g., eq.~(3) of the main text). To update the environment in a semi-smooth fashion (even if the next mutation has yet to occur), we include an ``environment update'' event that occurs at a rate comparable to the typical timescale of a fixation event.

\section{\label{sec:level2}Constructing Environment Pairs}
In this work, we studied evolution of genotypes after switching from one environment to another: An initially random genotype was first computationally evolved in an environment $\vec{E}_1$ until no beneficial mutations remain (highly adapted to $\vec{E}_1$). We then either directly or slowly switch exposure to environment $\vec{E}_2$. Although there are many features of environment pairs that may matter for an evolving genotype, here, for simplicity, we focus on characterizing each environment pair $(\vec{E}_1, \vec{E}_2)$ by the Euclidean distance between them, $\Delta E = \norm{\vec{E}_2 - \vec{E}_1}$.

\begin{figure}
    \centering
    \includegraphics[width=10cm]{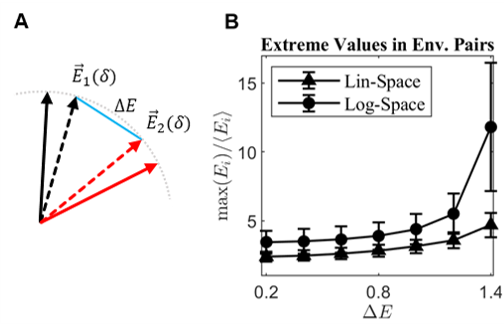}
    \caption{(A) Schematic of environment pair construction: Two $L$-dimensional vectors are randomly drawn and any negative entries are capped at 0 (red and black solid arrows). The two random vectors are then rotated (parameterized by eq.~\ref{eq:linSpacePairs}) until the desired $\Delta E$ is obtained (red and black dashed arrows). (B) Random environment pairs were constructed using either eq.~\ref{eq:linSpacePairs} or ~\ref{eq:logSpacePairs} over a range of $\Delta E$. For each environment, the maximum component is divided by the average of the components and plotted against $\Delta E$. Error bars correspond to $\pm 1$ standard deviation over 100 replicates. For $\Delta E > 1$, the log-space construction has extreme values that rapidly grow with increasing $\Delta E$, which is precisely the region of interest for this work.}
    \label{fig:S2}
\end{figure}

To construct a random pair with specified $\Delta E$, we generate two $L$-dimensional random vectors $(\vec{E}_A,\vec{E}_B)$ from a normal distribution with $\mu = 1, \sigma^2 = 1$ and rotate these vectors towards or away from each other (Fig.~\ref{fig:S2}A), similar to the approach of Ref.~\cite{tikhonov2020}. Specifically, the desired $\Delta E$ is attained by rotating the two random vectors away from their arithmetic mean $\vec{E} \equiv \frac{\vec{E}_A +\vec{E}_B}{2}$,  according to the following parameterization:
\begin{equation} \label{eq:linSpacePairs}
\begin{aligned}
    \vec{E}_1(\delta) &= \mathrm{normalize}\left[\mathrm{max}\left(\vec{E} + \frac{\delta}{2}(\vec{E}_A - \vec{E}_B),0\right)\right] \\
    \vec{E}_2(\delta) &= \mathrm{normalize}\left[\mathrm{max}\left(\vec{E} - \frac{\delta}{2}(\vec{E}_A - \vec{E}_B),0\right)\right]
\end{aligned}
\end{equation}
where the ``normalize'' operation normalizes a vector to unit length and $\mathrm{max(...,0)}$ acts component-wise to ensure that each component is nonnegative. Eq.~\eqref{eq:linSpacePairs} thus parametrically defines a function $\Delta E(\delta)$ that can be inverted for obtaining a random pair of environments $(\vec E_1, \vec E_2)$ a given $\Delta E$ apart. By construction, both vectors $\vec E_1$ and $\vec E_2$ obtained in this way have unit length and only nonnegative components.

Note that this is slightly different from the precise approach adopted in Ref.~\cite{tikhonov2020}. In that work, the rotation was performed in log-space: $\vec{E}_{1,2}(\delta) = \mathrm{normalize}[\vec{E}'_{1,2}(\delta)]$, where
\begin{equation} \label{eq:logSpacePairs}
\begin{aligned}
    \log \vec{E}'_1(\delta) &= \log\vec{E} + \frac{\delta}{2}(\log\vec{E}_A - \log\vec{E}_B) \\
    \log\vec{E}'_2(\delta) &= \log\vec{E} - \frac{\delta}{2}(\log\vec{E}_A - \log\vec{E}_B)
\end{aligned}
\end{equation}
and logarithms are applied component-wise. In Ref.~\cite{tikhonov2020}, the protocol Eq.~\eqref{eq:logSpacePairs} was adopted as the simplest approach that naturally preserved nonnegativity of vector entries, without the need for explicit truncation. However, for large $\Delta E$, environment pairs constructed in log-space will typically possess extremely large entries (see Fig.~\ref{fig:S2}B) that focus the majority of selection pressure on a few traits. Since much of our attention in this work concerns the large-$\Delta E$ regime (see, e.g., Figs.~4B,C in the main text), we opted for the linear-space construction of environment pairs, as defined in eq.~\ref{eq:linSpacePairs}.

\section{\label{sec:level3}Raw versus Gaussian Smoothed $s K_0$ and $s K_0/K$}
\begin{figure*}
    \centering
    \includegraphics[width=1\textwidth]{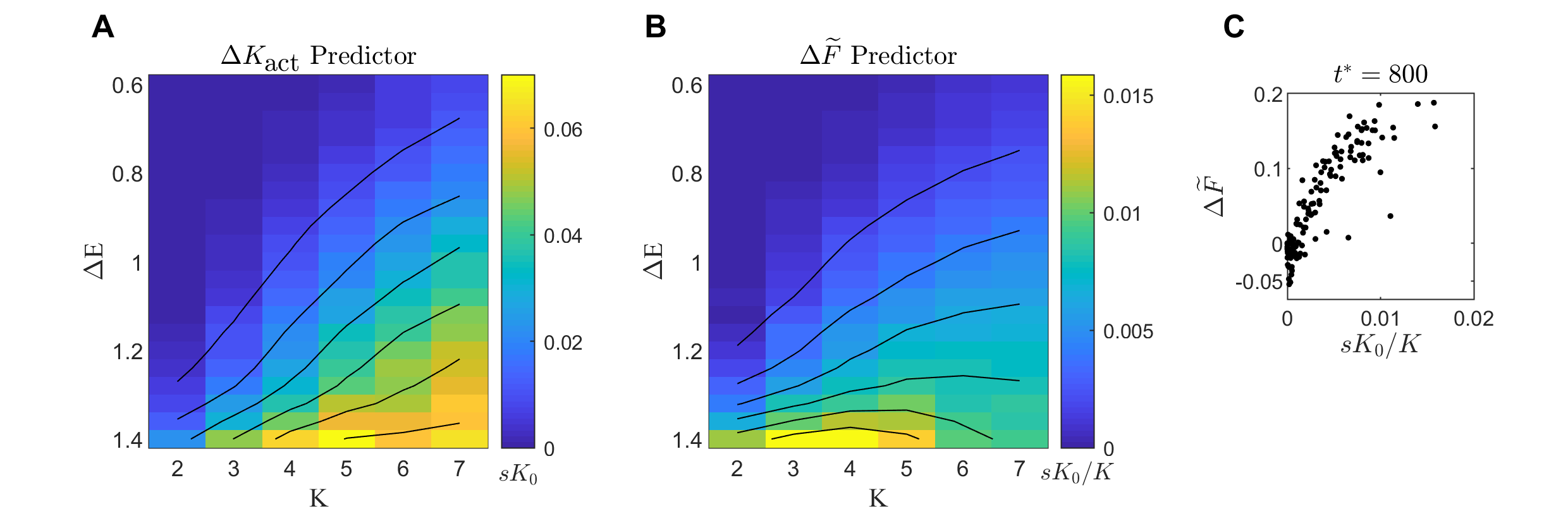}
    \caption{(A-B) Heatmap of feedback strength scaled by the number or fraction of inactive systems during early-times of evolution (first 3 mutations), $s K_0$ in A and $s K_0 / K s$ in B, with varying number of systems $K$ and degree of environment change $\Delta E$. The contour lines of Gaussian-smoothed $s K_0$ and $s K_0 / K s$ (filter width $\sigma = 1$) overlaid on top as done in Fig.~4B,C in main text. (C) Relative fitness gain $\Delta\widetilde{F}(t^*)$, as defined in eq. (4) in main text, measured at $t^* = 800$ (2-times later than the observation time in the main text) and scattered against the $s K_0 \ K$ data from (A). Each data point is an average over 300 trajectories. }
    \label{fig:S3}
\end{figure*}

Fig.~4B\&D traces a long-term evolutionary effect -- namely, the relative fitness gain $\Delta\widetilde{F}(t^*)$ that a slow exposure (SE) protocol provides compared to direct exposure (DE) to a novel environment -- to the early-time property of feedback loop strength, $s$. As described in the main text, the difference in number of active systems $\Delta K_{\act}$ between SE and DE is predicted by $s K_0$, where $K_0$ is the number of inactive systems at $t=0$ (the time at which the environment switches from $\vec{E}_1$ to $\vec{E}_2$ for the DE protocol). In turn, we reasoned $\Delta\widetilde{F}(t^*)\sim s K_0 / K$ for sufficiently late observation time $t^*$. Figs.~\ref{fig:S3}A,~B provide heatmaps of the raw $s K_0$ and $s K_0 / K$ values, respectively, for each $(K,\Delta E)$ parameter combination (average over 300 trajectories), from which the contour lines used in Fig.~4B,C were obtained after smoothing with a Gaussian kernel (of width equivalent to 1 heatmap pixel).

\section{\label{sec:level4}Dependence of Fitness Improvement on Observation Time}
The $\Delta\widetilde{F}(t^*)$ reported in panels B and E of Fig.~4 in the main text were measured at $t^* = 400$. Fig.~\ref{fig:S3}C replots the same results for $t^*=800$, demonstrating that the observation of Fig.~4 in the main text is not sensitive to the particular choice of the late-time observation point.

\end{document}